\documentclass{beamer}\mode<presentation>{\usetheme{Madrid}}
\setbeamercovered{highly dynamic}
\usepackage{graphicx}
\usepackage{amssymb,amsmath}
\newcommand{\lb}{\label}
\newcommand{\bfi}{\begin{figure}}
\newcommand{\efi}{\end{figure}}

\newcounter{saveenumi}
\newcommand{\seti}{\setcounter{saveenumi}{\value{enumi}}}
\newcommand{\conti}{\setcounter{enumi}{\value{saveenumi}}}

\newcommand{\be}{$$}
\newcommand{\ee}{$$}

\begin{document}
\title[inciple, C.D.~Froggatt, C.R.~Das, L.V.Laperashvili]
{Gravi-Weak Unification and Multiple Point Principle}
\author[\,Gravi-weak Unification and Multiple Point Principle]{\small
\textbf{C.D.~Froggatt} \\
Glasgow University, Glasgow, Scotland\\[1.5mm]
\textbf{C.R.~Das} \\
CFTP, Technical University of Lisbon, Portugal\\
and\\
University of Jyvaskyla, Finland\\[1.5mm]
\textbf{\alert{L.V.~Laperashvili}} \\
ITEP, Moscow, Russia\\[1.5mm]
\textbf{H.B.~Nielsen} \\
The Niels Bohr Institute, Copenhagen, Denmark\\[1.5mm]
\textbf{A.~Tureanu} \\
University of Helsinki, Finland\\[4mm]
{\color{blue}\bf
          International Conference of RAS\\
       Protvino, November 5-8, 2013}}
\date[hvili,H.B.~Nielsen, A.~Tureanu]{}


\frame{\titlepage}


\section*{Outline}
\begin{frame}
\frametitle{Outline} \tableofcontents
\end{frame}


\section{Introduction}


\begin{frame}
\frametitle{Introduction}
We construct a model unifying gravity with some, e.g. weak,
$SU(2)$ gauge and the ``Higgs" scalar fields.\\[1mm]

We assume the existence of a visible and an invisible (hidden)
sector of the Universe. The hidden world is a Mirror World which
is not identical with the visible one.\\[1mm]

This Hidden World (HW) is assumed
to be a Mirror World (MW) with broken mirror parity (MP).\\[1mm]

We used the extension of Plebanski's 4-dimensional gravitational
theory, in which the fundamental fields are two-forms containing
tetrads and spin connections, and in addition certain auxiliary
fields.\\[1mm]

We develop the model of Unification of gravity, gauge and Higgs
fields suggested in the reference:

{\color{magenta}\bf LSS model}:

\end{frame}


\section{Unification models and the Plebanski's theory of gravity}


\begin{frame}
\frametitle{Unification models and the Plebanski's theory of gravity}
Originally General Relativity (GR) was formulated by Einstein as
the dynamics of a metric, $g_{\mu\nu}$. Later Plebanski, Ashtekar
and others  presented GR in a self-dual approach:

\end{frame}


\begin{frame}
\frametitle{Unification models and the Plebanski's theory of gravity}
In this approach, the true configuration variable is a connection
corresponding to the gauging of the local Lorentz group,
$SO(1,3)$, and the spin group, $Spin(1,3)$.\\[2mm]

In general, in the unification models, the fundamental variable is
a connection, $A$, valued in a Lie algebra, $\mathfrak g$, that
includes a subalgebra $\tilde {\mathfrak  g}$:
\be\tilde {\mathfrak  g} = {\mathfrak  g}^{(spacetime)}
\oplus
{\mathfrak g}_{YM},\lb{1} \ee
which is the direct sum of the Lorentz algebra and a Yang--Mills
gauge algebra.
\end{frame}


\begin{frame}
\frametitle{Unification models and the Plebanski's theory of gravity}
In {\color{magenta}\bf LSS model}:\\[2mm]

${\mathfrak g}^{(spacetime)}= \mathfrak{ spin}(1,3)$ is the
gravitational gauge algebra, and the Yang--Mills gauge algebra is
a spin algebra, ${\mathfrak g}_{YM} = \mathfrak {spin}(N)$.
Finally, a model of unification of gravity, the $SU(N)$, or
$SO(N)$, gauge fields and Higgs bosons is based on the full
initial gauge algebra of type $\mathfrak g = \mathfrak
{spin}(p,q)$.\\[2mm]

In the Plebanski's formulation of the 4-dimensional theory of
gravity the gravitational action is the product of two 2-forms,
which are constructed from the connections $A^{IJ}$ and tetrads
(or frames) $e^I$ considered as independent dynamical variables.
\end{frame}


\begin{frame}
\frametitle{Unification models and the Plebanski's theory of gravity}
Both $A^{IJ}$ and $e^I$, also $A$, are 1-forms:
\be   A^{IJ} =  A_{\mu}^{IJ}dx^{\mu} \quad {\mbox{and}}\quad
       e^I = e_{\mu}^Idx^{\mu},
                          \lb{2} \ee
\be  A = \frac 12 A^{IJ}\gamma_{IJ}. \lb{3} \ee
Here the bivector generators $\gamma_{IJ}$  can be understood as
the product of $Cl(1,3)$ Clifford algebra basis vectors:
$$\gamma_{IJ}=\gamma_{I}\gamma_{J}.$$

The indices $I,J = 0,1,2,3$ refer to the spacetime with Minkowski
metric $$\eta_{IJ}: \eta^{IJ} = {\rm diag}(1,-1,-1,-1).$$ This
is a flat space which is tangential to the curved space with the
metric $g_{\mu\nu}$.
\end{frame}


\begin{frame}
\frametitle{Unification models and the Plebanski's theory of gravity}
The world interval is represented as $$ds^2 = \eta_{IJ}e^I \otimes
e^J,$$ i.e.
\be g_{\mu\nu} = \eta_{IJ} e^I_{\mu}\otimes e^J_{\nu}.
 \lb{4} \ee
Considering the case of the Minkowski flat spacetime with the
group of symmetry $SO(1,3)$, we have the capital latin indices
$I,J,...=0,1,2,3$.\\[2mm]

In the general case of the gauge symmetry $\mathfrak G$ with the Lie
algebra $\mathfrak g = spin(p,q)$, we have $I,J = 1,2,...,p+q$.\\[2mm]

The 2-forms $B^{IJ}$ and $F^{IJ}$ are defined as:
\be
      B^{IJ} = e^I\wedge e^J = \frac 12
      e_{\mu}^Ie_{\nu}^Jdx^{\mu}\wedge dx^{\nu},
                    \lb{5} \ee
\be
      F^{IJ} = \frac 12 F_{\mu\nu}^{IJ}dx^{\mu}\wedge dx^{\nu}.
                    \lb{6} \ee
\end{frame}


\begin{frame}
\frametitle{Unification models and the Plebanski's theory of gravity}
Here the tensor $F_{\mu\nu}^{IJ}$ is the field strength of the
spin connection $A_{\mu}^{IJ}$:
\be
     F_{\mu\nu}^{IJ} = \partial_{\mu}A_{\nu}^{IJ} -
         \partial_{\nu}A_{\mu}^{IJ} - \left[A_{\mu}, A_{\nu}\right]^{IJ},
                               \lb{7} \ee
which determines the Riemann--Cartan curvature:
\be
       R_{\kappa \lambda \mu \nu} = e_{\kappa}^I e_{\lambda}^JF_{\mu\nu}^{IJ}.
                  \lb{8} \ee
We also consider the 2-forms $B$ and $F$:
\be B= \frac 12 B^{IJ}\gamma_{IJ} \quad  {\rm and} \quad  F= \frac
12 F^{IJ}\gamma_{IJ},  \lb{9} \ee
\be F = dA + \frac 12 \left[A, A\right].  \lb{10} \ee
\end{frame}


\begin{frame}
\frametitle{Unification models and the Plebanski's theory of gravity}
In the Plebanski's BF-theory, the gravitational action with
nonzero cosmological constant $\Lambda$ is given by the integral:
\be  I_{GR} = \frac{1}{\kappa^2}\int
\epsilon^{IJKL}\left(B^{IJ}\wedge
    F^{KL} + \frac{\Lambda}{4}B^{IJ}\wedge B^{KL}\right),
                                  \lb{11} \ee
where $\kappa^2=8\pi G_N$, $G_N$ is the gravitational constant, $
M_{Pl}^{red.} = 1/{\sqrt{8\pi G_N}}$.\\[2mm]

For any antisymmetric tensors $F_{\mu\nu}$ there exist dual
tensors given by the Hodge star dual operation:
\be
 F^*_{\mu\nu}\equiv \frac {1}{2\sqrt{-g}}\epsilon
 ^{\rho\sigma}_{\mu\nu}F_{\rho\sigma}.
                                 \lb{12} \ee
\end{frame}


\begin{frame}
\frametitle{Unification models and the Plebanski's theory of gravity}
For any antisymmetric tensors $A^{IJ}$ there exists dual
operation:
\be  A^{\star IJ} = \frac 12 \epsilon^{IJKL}A^{KL}.
                                 \lb{13} \ee
Here $\epsilon$ is the completely antisymmetric tensor with
$\epsilon^{0123} = 1$.\\[2mm]

We can define the algebraic self-dual ($+$) and anti-self-dual ($-$)
components of $A^{IJ}$:
\be A^{(\pm)\,IJ}=\left({\cal P}^{\pm}A\right)^{IJ} = \frac 12 \left(A^{IJ} \pm
iA^{\star\,IJ}\right).
                                 \lb{14} \ee
\end{frame}


\begin{frame}
\frametitle{Unification models and the Plebanski's theory of gravity}
The two projectors ${\cal P}^{\pm}= \frac 12(\delta^{IJ}_{KL} \pm
\epsilon^{IJ}_{KL})$ realize explicitly the familiar homomorphism:
\be
   \mathfrak{so}(1,3) = \mathfrak{su}(2)_R \oplus
   \mathfrak{su}(2)_L,
                                 \lb{15} \ee
which rather than self-dual and anti-self-dual are more commonly
dubbed right-handed and left-handed.\\[2mm]

Then we define
\be
                 A^{(\pm)i} = A^{(\pm) 0i},
                     \lb{16} \ee
with $\large i = 1,2,3$ as an adjoint index of $SU(2)_L^{(grav)}$.
\end{frame}


\begin{frame}
\frametitle{Unification models and the Plebanski's theory of gravity}
The correct gauge was chosen by Plebanski, when he introduced in
the gravitational action the Lagrange multipliers $\psi_{ij}$ --
an auxiliary fields, symmetric and traceless. These auxiliary
fields $\psi_{ij}$ provide the correct number of constraints.\\[2mm]

Including the constraints, we obtain the following gravitational
action:
\be I(\Sigma,A,\psi) = \frac{1}{\kappa^2} \int \left[\Sigma^i\wedge F^i
+
 \left(\Psi^{-1}\right)_{ij}\Sigma^i\wedge \Sigma^j\right].
                      \lb{18} \ee
The usual notations $\Sigma^i=2B^{0i}$ and:
\be   \left(\Psi^{-1}\right)_{ij} = \psi_{ij} - \frac{\Lambda}{6}\delta_{ij}.
                     \lb{19} \ee
Then Plebanski and other authors  suggested to consider in the
visible sector of our Universe the left-handed
$\mathfrak{su}(2)_L$-invariant gravitational action  with
self-dual $F=F^{(+)i}$ and $\Sigma=\Sigma^{(+)i}$, which is
equivalent to the Einstein-Hilbert gravity.
\end{frame}


\section{Graviweak action in the visible sector of the Universe}


\begin{frame}
\frametitle{Graviweak action in the visible sector of the Universe}
In Ref.:

\vspace{2mm}
developing the graviweak unification model in the visible sector
of the Universe, we started with a $\mathfrak g = \mathfrak
{spin}(4,4)$-invariant extended Plebanski's action:
\be I(A, B, \Phi) = \frac{1}{g_{uni}} \int_{\mathfrak M}\left\langle BF
+  \frac{\Lambda_0}{4}BB + B\Phi B + \frac 13 B\Phi \Phi \Phi B
\right\rangle. \lb{22} \ee
The wedge product $\langle...\rangle$ is assumed between the
forms.\\[2mm]

In this action with a parameter of the unification $g_{uni}$, the
connection, $A = A^{IJ}\gamma_{IJ}$, is the independent physical
variable describing the geometry of the spacetime, while $\Phi$,
or $\Phi_{IJKL}$, are auxiliary fields.
\end{frame}


\begin{frame}
\frametitle{Graviweak action in the visible sector of the Universe}
Here, $$ F = dA + \frac 12 [A, A]$$ is the curvature and
$$B=B^{IJ}\gamma_{IJ}$$ is a $\mathfrak{spin}(4,4)$-valued 2-form
auxiliary field. The generators
$$\gamma_{IJ}=\gamma_{I}\gamma_{J}$$ of the
$\mathfrak{spin}(4,4)$-group have indices running over all
$8\times 8$ values: $I,J = 1,2,...,7,8$\\[2mm] ($I,J=1,5,6,7$ - time like
components, and $I,J=2,3,4,8$ - spatial ones).\\[2mm]

$\Lambda_0$ is the Plebanski's bare cosmological constant.
\end{frame}


\begin{frame}
\frametitle{Graviweak action in the visible sector of the Universe}
Varying the fields $A,B$ and $\Phi$, we obtained the field
equations:
\be  {\cal D}B = dB + [A,B] = 0,  \lb{23} \ee
where ${\cal D}$ is the covariant derivative, $${\cal
D}_{\mu}^{IJ} = \delta^{IJ}\partial_{\mu} - A_{\mu}^{IJ},$$
and
\be   F = -2\left(\Phi +\frac 13\Phi \Phi \Phi\right)B,  \lb{24}
\ee
\be    B^{IJ} B^{KL} = - \frac 1{16} B^{IJ}
\Phi^{KL}_{MN}\Phi^{MN}_{PQ}B^{PQ}. \lb{25} \ee
The first equation describes the dynamics, while
last two determine $B$ and $\Phi$ respectively.\\[2mm]

Here we assumed that $\Lambda_0=0$.
\end{frame}


\begin{frame}
\frametitle{Graviweak action in the visible sector of the Universe}
With help of the equations of motion, we have obtained the
following $\mathfrak g$-invariant gravitational, and weak
$SU(2)_L$ gauge and Higgs fields action:
\be
   I(e,A) = \frac{3}{8g_{uni}}\int_{\mathfrak M} \left\langle F F^{\star} \right\rangle.
   \lb{26} \ee
Considering the spontaneous symmetry breaking of the above $\mathfrak
g$-invariant action we see that it produces the
dynamics of the $ SU(2)_L$-gravity, and the $SU(2)_L$ gauge and
Higgs fields with subalgebra
$$  \tilde {\mathfrak g} = {{\mathfrak su}(2)}^{(grav)}_L
\oplus {\mathfrak su}(2)_L.$$
The indices $\large a, b
\in\{0,1,2,3\}$ are used to sum over a subset of $I, J \in {1,2,
...,7,8}$ for $I,J=1,2,3,4$, and thereby select a $\mathfrak
{spin}(1,3)$ subalgebra of $\mathfrak {spin}(4,4)$. The indices $
m, n \in \{5,6,7,8\}$ sum over the rest. We also consider $\large
i, j \in \{1,2,3\}$, thus selecting a $\mathfrak {su}(2)_L^{grav}$
subalgebra of $\mathfrak {spin}(4,4)_L$.
\end{frame}


\begin{frame}
\frametitle{Graviweak action in the visible sector of the Universe}
The  spontaneous symmetry breaking (SSB) of the graviweak
unification gives separate parts of the connection in terms of the
following 2-forms:
\be   A = \frac 12 \omega + \frac 14 E + A_W  \lb{28} \ee
with the gravitational spin connection:
\be   \omega = \omega^{ab}\gamma_{ab}, \lb{33} \ee
or
\be  \omega =  \omega^i \sigma_i, \lb{34} \ee
which is valued in $\mathfrak {su}(2)_L^{(grav)}$. Here $\sigma_i$
are Pauli matrices, $i=1,2,3$.
\end{frame}


\begin{frame}
\frametitle{Graviweak action in the visible sector of the Universe}
The frame-Higgs connection
\be   E = E^{am}\gamma_{am}   \lb{35} \ee
is valued in the off-diagonal complement of $\mathfrak
{spin}(4,4)$, and assumed to have the expression:
\be   E = e\phi = e^a_{\mu}\sigma_a\phi^i\tau_i dx^{\mu},
                                          \lb{36} \ee
what corresponds to $\mathfrak {su}(2)$-subgroup of the Clifford
algebra, where:
\be   E = e\phi = e^a_{\mu}\gamma_a\phi^m\gamma_m dx^{\mu}.
                                          \lb{36a} \ee
The field $\phi=\phi^i\tau_i$ is the scalar Higgs of $\mathfrak
su(2)_L$,\\[2mm]

$\tau_i$ are Pauli matrices, $i=1,2,3$.
\end{frame}


\begin{frame}
\frametitle{Graviweak action in the visible sector of the Universe}
The gauge field:
\be  A_W = \frac 12 A^{mn}\gamma_{mn}, \lb{37} \ee
or
\be   A_W = \frac 12 A_W^i \tau_i, \lb{38} \ee
is valued in $\mathfrak su(2)_L$.\\[2mm]

\begin{exampleblock}{Finally, we have the following graviweak action for the
gravitational, $SU(2)_L$ gauge and Higgs fields in the ordinary
(visible) sector of the Universe:}
$$  I_{GWU}\left(e,\phi,A,A_W\right)= - \frac{3}{8g_{uni}}
\int_{\mathfrak M} d^4x|e|\left(-\frac 1{16}{|\phi|}^2 R +
\frac{3}{32}{|\phi|}^4\right. $$
\be \left.- \frac 1{16}{R_{ab}}^{cd} {R^{ab}}_{cd}+ \frac 12 {\cal D}_a{\phi^{\dagger}}
{\cal D}^a\phi + \frac 14 {F_W^i}_{ab}{F_W^i}^{ab} \right). \lb{51} \ee
\end{exampleblock}
\end{frame}


\begin{frame}
\frametitle{Graviweak action in the visible sector of the Universe}
Here $R$ is the Riemann curvature scalar,  $|\phi|^2 =
{\phi}^{\dag}\phi$ is the squared magnitude of the Higgs field,
${\cal D} \phi = d\phi + [A_W,\phi]$ is the covariant derivative
of the Higgs field, and \\ $F_W = dA_W + [A_W,A_W]$ is the
curvature of the gauge field $A_W$.\\[2mm]

The third term of the action is a Stephenson-Kilmister-Yang (SKY)
modification to the standard gravitation related to the
Gauss-Bonnet topological action:

\end{frame}


\begin{frame}
\frametitle{Graviweak action in the visible sector of the Universe}
The nontrivial vacuum solutions to the action give the
non-vanishing Higgs vacuum expectation value (VEV):
$$v=\langle\phi\rangle=\phi_0,$$ at which the standard Higgs
potential has an extremum corresponding to the de Sitter spacetime
background solution: $$v^2 = \frac{R_0}{3}.$$ Here $R_0 > 0$ is a
constant background scalar curvature.
\end{frame}


\begin{frame}
\frametitle{Graviweak action in the visible sector of the Universe}
After the symmetry breaking of the graviweak unification, we
obtain:

\begin{enumerate}
\item the {Newton'}s constant in our Universe, equal to
\be 16\pi G_N = \frac{128g_{uni}}{3v^2}; \lb{52} \ee
\item the  cosmological constant $\Lambda$:
 \be \Lambda = \frac 34 v^2;  \lb{53} \ee
\item the weak coupling constant:
\be   g_W^2 = \frac{8g_{uni}}{3}.  \lb{54} \ee
\end{enumerate}
\end{frame}


\begin{frame}
\frametitle{Graviweak action in the visible sector of the Universe}
All physical constants are determined by a parameter $g_{uni}$ and
the Higgs VEV $v$. It is necessary to note that all parameters --
Newton's constant, the cosmological constant, the gauge couplings
$g_{YM}=g_W$, considered in the action -- are bare parameters,
which refer to the Planck scale.

\begin{exampleblock}{}
Assuming that the scalar field $\phi$ is usual Higgs doublet
of the SM,
\end{exampleblock}

we can use the experimentally known value of $G_N$, where
$$M_{Pl}^{red.} = 1/{\sqrt{8\pi G_N}}\approx 2.43\times
10^{18}\,\,{\rm GeV},$$ and obtain the value of $g_{uni}$, if we
relate the value $g^2_W=8g_{uni}/3$ with the value of $g_2^2$
obtained by the extrapolation of experimental values of running
$\alpha_2=g_2^2/4\pi $ from the Electroweak scale to the Planck
scale.

\end{frame}


\begin{frame}
\frametitle{Graviweak action in the visible sector of the
Universe}

See:

\end{frame}


\begin{frame}
\frametitle{Graviweak action in the visible sector of the Universe}
\begin{block}{Fig.~1: The evolution of the inverse SM  fine
structure constants as functions of $x$ ($x = \log_{10}\mu \,({\rm GeV})$)
up to the Planck scale $M_{Pl}$.}
\bfi\centering
\includegraphics[height=93mm,keepaspectratio=true,angle=-90]
{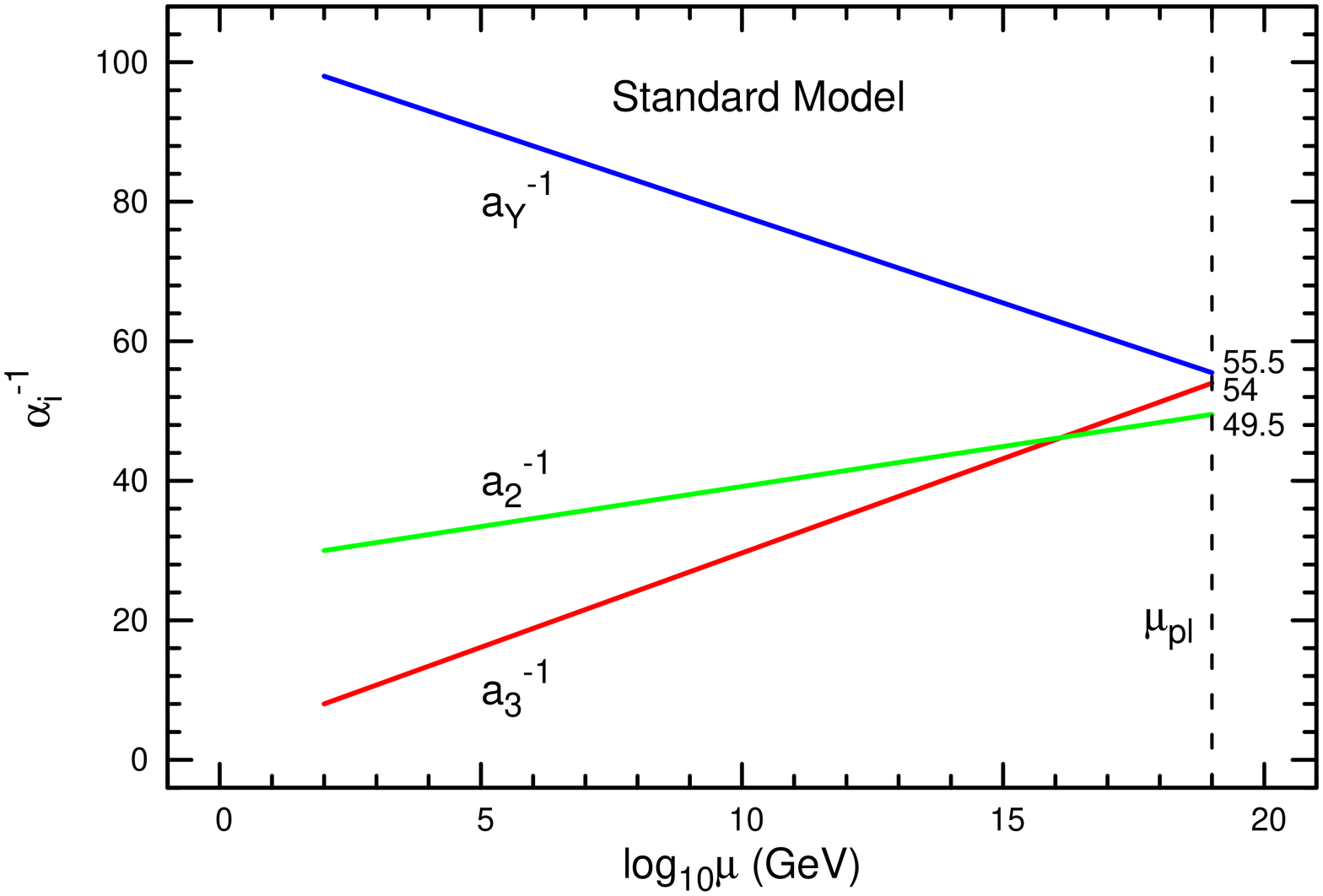}
\efi
\end{block}
\end{frame}


\begin{frame}
\frametitle{Graviweak action in the visible sector of the Universe}
Then we have no agreement of the graviweak unification with
experimental measurements, giving $$g_{uni}\sim 0.1,$$ if we have
the well-known Newtonian constant $G_N$.\\[2mm]

In this case we have:
$$
 v\approx 246 \,\,{\rm GeV}, \quad g_{uni}\approx 5\cdot 10^{-34}\sim 0.$$
However, the self-consistent graviweak unification can be obtained
in the model with the \alert{Multiple Point Principle} developed by
H.B.Nielsen and his collaborators (D.L.~Bennett, C.D.~ Froggatt,
R.B.~Nevzorov, L.V.L., etc.).
\end{frame}


\section{Multiple Point Principle}


\begin{frame}
\frametitle{Multiple Point Principle}
The vast majority of the available experimental information is
already explained by the SM. All accelerator physics is in
agreement with the SM, except for neutrino oscillations.\\[2mm]

\begin{exampleblock}{}
One of the main goals of physics today is to find out the fundamental
theory beyond the SM. In first approximation we might ignore the
indications of new physics and consider the possibility that the
SM essentially represents physics well up to the Planck scale.\\[2mm]
\end{exampleblock}
\end{frame}


\begin{frame}
\frametitle{Multiple Point Principle}
In Ref. {\color{magenta}\bf FLN}:

we suggested a scenario, using only the pure SM, in which an
exponentially huge ratio between the fundamental (Planck) and
Electroweak scales results:
\be
     \frac{\mu_{fund}}{\mu_{ew}} \sim e^{40}\sim 10^{17}. \lb{A}
\ee
We suggest a model, which contains simply the SM itself up to the
scale $\sim 10^{19}$ {\rm GeV}, or at least $10^{16}$ {\rm GeV}.

\begin{exampleblock}{}
This model reminds us ``the Bjorken-Rosner nightmare'', when there
is no new physics up to the Planck scale (except perhaps see-saw
scale). In such a scenario it is reasonable to assume the
existence of a simple postulate, which helps us to explain the SM
parameters: couplings, masses and mixing angles.
\end{exampleblock}
\end{frame}


\begin{frame}
\frametitle{Multiple Point Principle}
\begin{exampleblock}{Such a postulate is based on a phenomenologically required
result in cosmology (see Particle Data Group):} the cosmological constant
is zero, or approximately zero, meaning that the vacuum energy
density (dark energy) is very small. A priori it is quite possible
for a quantum field theory to have several minima of the effective
potential as a function of its scalar field. Postulating zero
cosmological constant, we can assume that all the vacua, which
might exist in Nature (as minima of the effective potential),
should have zero, or approximately zero cosmological constant.
This postulate corresponds to what we call the Multiple Point
Principle (MPP).
\end{exampleblock}
\end{frame}


\begin{frame}
\frametitle{Multiple Point Principle}
\begin{alertblock}{MPP postulates:}
 {\it There are many vacua with the same energy density, or cosmological
constant, and all cosmological constants are zero or approximately zero.}
\end{alertblock}

If we have the first minimum  of the Higgs effective potential at
$v = \phi_{min1}\approx 246 \,\, {\rm GeV}$,  then we have the
possible existence of a second (non-standard) minimum at the
fundamental scale:
\be
            \phi_{min2} \gg v =  \phi_{min1}.      \lb{51mp}
\ee
In accord with cosmological results, we take the cosmological
constants $C$ for both vacua equal to zero (or approximately
zero): $C=0$ (or $C\approx 0$).
\end{frame}


\begin{frame}
\frametitle{Multiple Point Principle}
The following requirements must be satisfied in order that the SM
effective potential should have two degenerate minima: \be
        V_{eff}(\phi_{min1}^2) = V_{eff}(\phi_{min2}^2) = 0,
        \lb{53mp}
\ee \be
        V'_{eff}(\phi_{min1}^2) = V'_{eff}(\phi_{min2}^2) = 0,
          \lb{54mp}
\ee where \be
         V'(\phi^2) = \frac{\partial V}{\partial \phi^2}.
                                             \lb{55mp}
\ee
These degeneracy conditions correspond to the MPP expectation.
\begin{alertblock}{}
{\it The first minimum is the
standard ``Weak scale minimum", and the second one is the
non-standard ``Fundamental scale minimum" (if it exists).}
\end{alertblock}

An illustrative schematic picture of $V_{eff}$ is presented in
Fig.~2.
\end{frame}


\begin{frame}
\frametitle{Multiple Point Principle}
\begin{block}{Fig.~2:}
\bfi\centering\includegraphics[height=70mm,keepaspectratio=true,angle=0]
{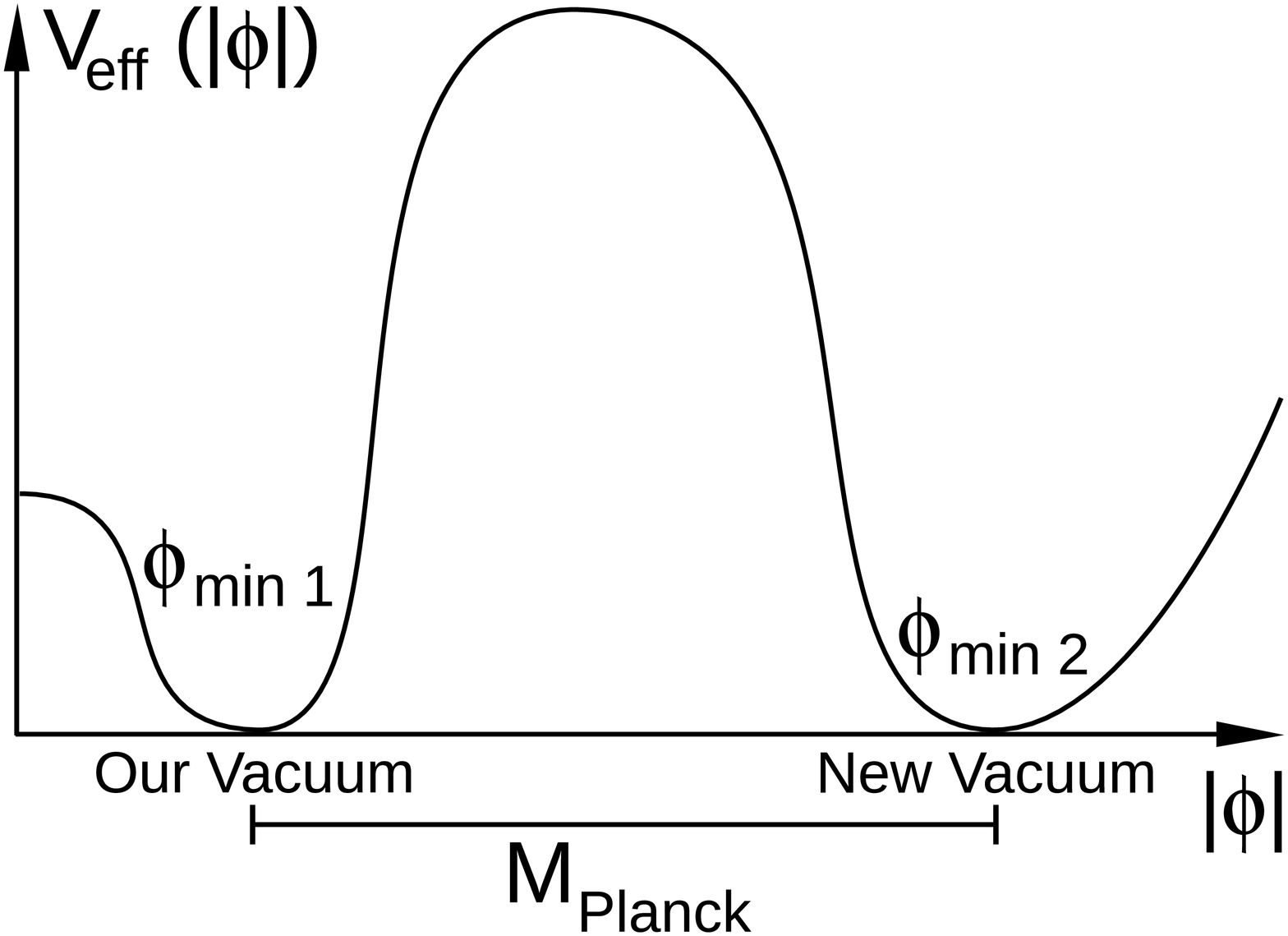}\efi
\end{block}
\end{frame}


\begin{frame}
\frametitle{Multiple Point Principle}
The predictions of Ref.:

\vspace{2mm}
for the top-quark and Higgs masses from the MPP requirement of the
existence of a second degenerate vacuum, were as follows:
\be M_t = 173 \pm 5\,\,{\rm GeV},\quad M_H = 135 \pm
9\,\,{\rm GeV}. \lb{56mp} \ee
For large values of the Higgs field: $\phi^2 \gg m^2$ the effective
potential $V_{eff}$ is very well approximated by the quartic term
$\lambda(\phi^{+}\phi)^2$ and the degeneracy conditions  give:
\be
       \lambda(\phi_{min2}) = 0, \lb{57mp}
\ee
\be
 \lambda'(\phi_{min2}) = 0.    \lb{58mp}
\ee
\end{frame}


\begin{frame}
\frametitle{Multiple Point Principle}
These conditions can be expressed in the form:
\be
      \beta_{\lambda}(\phi_{min2}, \lambda=0) = 0.  \lb{59mp}
\ee
In Ref. {\color{magenta}\bf FLN}:
the scale $v_2=\phi_{min2}$ (depending on the experimental data
uncertainties) was calculated in the second loop approximation of
the Higgs effective potential.\\[2mm]

It was shown that in the 2-loop approximation, the experimental
values of the coupling constants allow the existence of the SM
effective potential second minimum.\\[2mm]

We have calculated this position numerically. It exists in the
interval:
\be \phi_{min2}\in (10^{16}; 10^{22})\,\, {\rm GeV}.    \lb{74r} \ee
\end{frame}


\begin{frame}
\frametitle{Multiple Point Principle}
It is necessary to emphasize that, for the central values of the
experimental parameters, we predict the fundamental scale (the
position of the second minimum) to be close to the Planck scale
$\mu_{fundamental}\sim 10^{19}$ {\rm GeV}.\\[2mm]

The MPP-model of the SM also predicts a value for the Higgs mass.
In spite of the large uncertainty in the position of the second
minimum, the value of $\lambda$ at the Electroweak scale lies in
the narrow interval:
\be \lambda(M_t) \in (0.26; 0.34),     \lb{75r} \ee
which corresponds to the prediction of the interval of the Higgs
mass equal to:
\be  M_H \in (125; 143)\,\,{\rm GeV}.
                                       \lb{76r} \ee
This interval covers the LHC prediction of the Higgs mass
($\approx 126$ {\rm GeV}) and the prediction of Froggatt-Nielsen. ($M_H
= 135 \pm 9$ {\rm GeV}), what means that the radiative corrections to
the Higgs effective potential explains the value of the Higgs mass
observed at the LHC.
\end{frame}


\begin{frame}
\frametitle{Multiple Point Principle}
The same subject was considered in Ref.:

\vspace{2mm}
They obtained the result: $$ M_H = 129 \pm 2\,\,{\rm{GeV}},$$
which is very close to the result observed at the LHC: $M_H\approx 126.4$
{\rm{GeV}}. 
\end{frame}


\section{Planck scale values of the graviweak unification parameters}


\begin{frame}
\frametitle{Planck scale values of the graviweak unification
parameters} The SM-MPP model gives the self-consistent description
of the graviweak unification, giving the well-known value of $G_N$
with
$$
v_2\approx 2.5\times 10^{19}\,\,{\rm GeV},$$ and $$ g_{uni}\sim
0.1.$$

But the idea that the vacuum value of $\langle\phi\rangle$ could
be (according to the \alert{Multiple Point Principle}) the second
minimum $v_2$ of the Higgs field effective potential, turns out
not to be viable. The reason is that the Newtonian constant $G_N$
was measured in the vacuum $\langle\phi\rangle=v_1\approx 246$
{\rm GeV} - in the vacuum, in which we live.

Then other assumption can give a self-consistent description of
the graviweak unification: to consider the existence of other
scalar "Higgs" field $\varphi_{\theta}$, maybe existing only in
the Hidden sector of our Universe, having other $SU(2)'_\theta$
gauge and "Higgs" fields:

\end{frame}


\begin{frame}
\frametitle{Planck scale values of the graviweak unification
parameters}

See analogous example in Refs.:

\end{frame}


\begin{frame}
\frametitle{Planck scale values of the graviweak unification
parameters}

If there exists an axial $U(1)_A$ global symmetry, which is
spontaneously broken at the scale $f_{\theta}\sim 10^{19}$ GeV by
a complex scalar field $\varphi_{\theta}$, then we have:
$$ \varphi_{\theta} = (f_\theta + \sigma) \exp(i a_\theta/f_\theta).
$$
The boson $a_{\theta}$ (imaginary part of the scalar field
$\varphi_{\theta}$) is an axion and could be identified with a
massless Nambu-Goldstone (NG) boson if the $U(1)_A$ symmetry is
not spontaneously broken. However, the spontaneous breaking of the
global $U(1)_A$ by $SU(2)'_{\theta}$ instantons inverts
$a_{\theta}$ into a pseudo Nambu-Goldstone (PNG) boson. This PNG
axion has a mass squared:
$$ m^2\sim {\Lambda'_{\theta}}^4/f^2_{\theta}. $$

\end{frame}


\begin{frame}
\frametitle{Planck scale values of the graviweak unification
parameters}

In such a model we have the self-consistent graviweak unification,
and the condensates of $SU(2)'_{\theta}$-fields gives:
$$
\rho_{DE} \approx (2.3 \times 10^{-3} \,\, {\rm eV})^4.
$$
Thus, the field $\varphi_{\theta}$ (doublet or triplet of
$SU(2)'_{\theta}$), that has a Planck scale expectation value,
could have a better chance of being the scalar field unified with
gravity.
\end{frame}


\section{Graviweak action in the hidden sector of the Universe}


\begin{frame}
\frametitle{Graviweak action in the hidden sector of the Universe}
In Ref.:

\vspace{2mm}
we suggested to describe the gravity in the visible Universe by
the self-dual left-handed Plebanski's gravitational action, while
the gravity in the invisible (hidden) Universe -- by the
anti-self-dual right-handed gravitational action:
\begin{exampleblock}{}
\be I^{(\prime)}_{(gravity)}\left(\Sigma^{(\prime)},A^{(\prime)},\psi^{(\prime)}\right) =
   \frac {1}{{\kappa^{(\prime)}}^2}\int\left[{\Sigma^{(\prime)}}^i\wedge {F^{(\prime)}}^i +
\left({\Psi^{(\prime)}}^{-1}\right)_{ij}{\Sigma^{(\prime)}}^i\wedge {\Sigma^{(\prime)}}^j\right],
                      \lb{21h} \ee
\end{exampleblock}
where the superscript `prime' denotes the M-
or hidden H-world. Here $\Sigma = A^{(+)},\Sigma^{(+)}$ are self-dual
(left-handed) fields in the OW, and $A',\Sigma' =
A^{(-)},\Sigma^{(-)}$ are anti-self-dual (right-handed) fields in
the MW.
\end{frame}


\begin{frame}
\frametitle{Graviweak action in the hidden sector of the Universe}
Developing these ideas, we consider the graviweak unification
model in both sectors of the Universe, visible and invisible.\\[2mm]

Now we distinguish the following two actions:

\begin{enumerate}
\item the $\mathfrak {spin}(4,4)_L$-invariant action $I_{left}(A, B,
\Phi)$ with self-dual left-handed fields \\ $A=A^{(+)}$,
$B=B^{(+)}$ and auxiliary fields $\Phi_{IJKL}$ -- in the ordinary
(visible) world OW, and
\item the $\mathfrak {spin}(4,4)_R$-invariant action $I_{right}(A',
B', \Phi')$ with anti-self-dual right-handed fields $A'=A^{(-)}$,
$B'=B^{(-)}$ and auxiliary fields $\Phi'_{IJKL}$ -- in the hidden
(invisible) world MW.
\end{enumerate}

Instead of fields $A,B,F$ and $\Phi$,  similar equations hold in
MW for $A',B',F'$ and $\Phi'$.
\end{frame}


\begin{frame}
\frametitle{Graviweak action in the hidden sector of the Universe}
For completeness, we briefly present the spontaneous symmetry
breaking of the $\mathfrak g$-invariant action that produces the
dynamics of the $ SU(2)_L$-gravity, and the $SU(2)_L$ gauge and
Higgs fields with subalgebra
$$  \tilde {\mathfrak g} = {{\mathfrak su}(2)}^{(grav)}_L
\oplus {\mathfrak su}(2)_L.$$

Analogous equations are valid in the MW with the initial
$\mathfrak {spin}(4,4)_R$-algebra, and with a subalgebra:
\be  \tilde {\mathfrak g}' = {{\mathfrak su}(2)'}^{(grav)}_R
\oplus {\mathfrak su}(2)'_R.      \lb{27h} \ee
\end{frame}


\begin{frame}
\frametitle{Graviweak action in the hidden sector of the Universe}
\begin{exampleblock}{After the spontaneous symmetry breaking of the graviweak
unification, we have the following actions for gravitational
gauge and Higgs fields in the ordinary $SU(2)_L$
and hidden $SU(2)'_R$ sectors of the Universe:}
$$  I^{(\prime)}\left(e^{(\prime)},\phi^{(\prime)},A^{(\prime)},A^{(\prime)}_{W^{(\prime)}}\right)=
- \frac{3}{8g_{uni}}
\int_{\mathfrak M} d^4x|e^{(\prime)}|\left(-\frac 1{16}{|\phi^{(\prime)}|}^2
R^{(\prime)} \right.$$ \be
\left. + \frac{3}{32}{|\phi^{(\prime)}|}^4 - \frac
1{16}{{R^{(\prime)}}_{ab}}^{cd} {{R^{(\prime)}}^{ab}}_{cd}
 + \frac 12
{\cal D}_a{\phi^{(\prime)}}^{\dagger} {\cal D}^a\phi^{(\prime)} + \frac 14
{F^{(\prime)}}^{i}_{W^{(\prime)},ab}{F^{(\prime)}}^{i,ab}_{W^{(\prime)}}\right).
\lb{51h} \ee
\end{exampleblock}
\end{frame}


\begin{frame}
\frametitle{Graviweak action in the hidden sector of the Universe}
Here $g_{uni}=g'_{uni}$, since we assume that this equality is a
consequence of the existence of the Grand Unification at the early
stage of the Universe, when the mirror parity was unbroken. In
the action $R^{(\prime)}$ are the Riemann curvature scalars.  Similar
notations $A',\phi',A'_{W'}$ were used instead of the fields
$A,\phi,A_{W}$. The nontrivial vacuum solutions to the actions
give the non-vanishing Higgs vacuum expectation values
(VEVs): $v^{(\prime)}=\langle\phi^{(\prime)}\rangle=\phi^{(\prime)}_0.$

After the symmetry breaking of graviweak unification in the hidden
sector of the Universe, we obtain:
\be 16\pi G'_N=\frac{128g_{uni}}{3{v'}^2},\quad \Lambda' = \frac 34
{v'}^2, \quad {g'}^2_W=\frac{8g_{uni}}{3}, \lb{54h} \ee
and we have the following relations between O- and M-parameters:
\be G'_N = \frac{G_N}{\zeta^2}, \quad
  \Lambda' = \zeta^2 \Lambda, \quad M_{Pl}' = \zeta M_{Pl},
     \quad       g'_W = g_W.  \lb{55h} \ee
All physical constants of the Universe are determined by a
parameter $g_{uni}$ and the Higgs VEVs $v,v'$.
\end{frame}

\section{Dark energy of the Universe}


\begin{frame}
\frametitle{Dark energy of the Universe}
\begin{exampleblock}{In our theory the dark energy density of the Universe is given by
the expression:}
\be  \rho_{DE} = \rho_{vac} = \frac{\Lambda_{eff}}{8\pi G_N}  +
\frac{\Lambda'_{eff}}{8\pi G'_N},
                        \lb{64de} \ee
where
\be   \frac{\Lambda^{(\prime)}_{eff}}{8\pi G^{(\prime)}_N} =
\frac{\Lambda^{(\prime)}+\Lambda_0^{(\prime)}}{8\pi G^{(\prime)}_N} +
\rho_{vac}^{(SM^{(\prime)})},
                        \lb{64ade} \ee
\end{exampleblock}
All quantum fluctuations of the matter (SM and SM') contribute to
the vacuum energy density $\rho_{vac}$ of the Universe.
\end{frame}


\begin{frame}
\frametitle{Dark energy of the Universe}
Astrophysical measurements (Particle Data Group) give:
\be  \rho_{DE} \approx 0.73 \rho_{tot} \approx (2.3\times
10^{-3}\,\,{\rm eV})^4.
                        \lb{65de} \ee
We considered that all matter quantum fluctuations in SM and SM'
are compensated by the contribution of cosmological constants
$\Lambda$ and $\Lambda'$:
$$
\frac{\Lambda^{(\prime)}}{8\pi G^{(\prime)}_N} +
\rho_{vac}^{(SM^{(\prime)})}=0.$$
Taking into account the Einstein's cosmological constants
$\Lambda_0^{(\prime)}$ (presumably connected with zero-point energies
of gravitational fields), we can describe $\rho_{DE}$ by the
following way:
\begin{exampleblock}{}
\be \rho_{DE} = \frac{\Lambda_0}{8\pi G_N}  +
\frac{\Lambda'_0}{8\pi G'_N} =  \frac{\Lambda_0}{4\pi G_N}\approx
(2.3\times 10^{-3}\,\,{\rm eV})^4.
                        \lb{74de} \ee
\end{exampleblock}
\end{frame}


\section{Conclusions}


\begin{frame}
\frametitle{Conclusions}
\begin{enumerate}
\item We constructed a model  unifying gravity with some, e.g. weak,
$SU(2)$ gauge and ``Higgs" scalar fields.
\item We assumed the existence of a visible and an invisible (hidden)
sector of the Universe. The hidden world is a Mirror World which
is not identical with the visible one.
\item We used the extension of Plebanski's 4-dimensional
gravitational theory, in which the fundamental fields are
two-forms containing tetrads and spin connections, and in addition
certain auxiliary fields.
\item Considering a $Spin(4,4)$ invariant extended Plebanski action,
we started with a diffeomorphism invariant theory of a gauge field
taking values in a Lie algebra $\mathfrak g$, which is broken
spontaneously to the direct sum of the space-time Lorentz algebra
and the Yang-Mills algebra. i.e.: $\tilde {\mathfrak g} =
{\mathfrak su}(2)^{(grav)}_L \oplus {\mathfrak su}(2)_L$  in the
ordinary world, and $\tilde {\mathfrak g}' = {{\mathfrak
su}(2)'}^{(grav)}_R \oplus {\mathfrak su}(2)'_R$ in the hidden
world.
\seti
\end{enumerate}
\end{frame}


\begin{frame}
\frametitle{Conclusions}
\begin{enumerate}
\conti \item We recovered the actions for gravity, $SU(2)$
Yang-Mills and ``Higgs" fields in both (visible and invisible)
sectors of the Universe. 
\item After symmetry breaking of this GW unification its
physical constants (Newton's constants, cosmological constants,
Yang-Mills couplings, etc.), are determined by a parameter
$g_{uni}$ of the GW unification and by the Higgs VEVs. 
\item It is
discussed that if this ``Higgs" $\phi$ coming in the graviweak
unification could be the Higgs of the Standard Model, then the
idea that the vacuum value of $\langle\phi\rangle$ could be,
according to the MPP, an extra (second) minimum of the Higgs field
effective potential, turns out not to be viable. Then other scalar
``Higgs" field $\varphi_{\theta}$, giving the inflaton and the axion
fields, has a Planck scale expectation value, and could have a
better chance of being the scalar field unified with gravity.
\end{enumerate}
\end{frame}

\section{References}


\begin{frame}
\frametitle{References}
\begin{enumerate}
\item
A.~Garrett~Lisi, L.~Smolin and S.~Speziale,\\
{\it Unification of
gravity, gauge fields, and Higgs bosons,}\\
J. Phys. A {\bf 43}, 445401 (2010), arXiv:1004.4866.
\item
J.~F.~Plebanski,\\ {\it On the separation of Einsteinian
substructures,}\\ J. Math. Phys. {\bf 18}, 2511
(1977).
\item A.~Ashtekar,\\ {\it New variables for classical
and quantum gravity,}\\ Phys. Rev. Lett. {\bf 57}, 2244 (1986);
Phys. Rev. D {\bf 36}, 1587 (1987).
\item
R.~Capovilla, T.~Jacobson, J.~Dell and L.~J.~Mason,\\
{\it Selfdual two forms and gravity,}\\ Class. Quant. Grav. {\bf
8}, 41 (1991).
\item
R.~Capovilla, T.~Jacobson and J.~Dell,\\ {\it A pure spin
connection formulation of gravity,}\\ Class. Quant. Grav. {\bf 8},
59 (1991).
\seti
\end{enumerate}
\end{frame}

\begin{frame}
\frametitle{References}
\begin{enumerate}
\conti
\item
C.R.~Das, L.V.~Laperashvili and A.~Tureanu,\\
{\it Graviweak unification, invisible Universe and dark energy,}\\
Int. J. Mod. Phys. A {\bf 28}, 1350085 (2013), arXiv:1304.3069.
\item
E.W.~Mielke,\\ {\it Einsteinian gravity from a topological action,}\\
Phys. Rev. D {\bf 77}, 084020 (2008), arXiv:0707.3466.
\item
 G.~de~Berredo-Peixoto and I.L.~Shapiro,\\
{\it Conformal quantum gravity with the Gauss-Bonnet term,}\\
Phys. Rev. D {\bf 70}, 044024 (2004), arXiv:hep-th/0307030.
\item
M.B.~Gaete and M.~Hassaine,\\ {\it Topological black holes for
Einstein-Gauss-Bonnet gravity with a nonminimal scalar field,}\\
arXiv:1308.3076.
\seti
\end{enumerate}
\end{frame}

\begin{frame}
\frametitle{References}
\begin{enumerate}
\conti
\item
D.L.~Bennett, L.V.~Laperashvili and H.B.~Nielsen,\\ {\it Relation
between finestructure constants at the Planck scale from multiple
point principle,}\\ in: Proceedings to the 9th Workshop on 'What
Comes Beyond the Standard Models?' Bled, Slovenia, July 16-27,
2006 (DMFA, Zaloznistvo, Ljubljana, 2006); arXiv:hep-ph/0612250.
\item
 D.L.~Bennett, L.V.~Laperashvili and H.B.~Nielsen,\\ {\it
Finestructure constants at the Planck scale from multiple point
principle,}\\ in: Proceedings to the 10th Workshop on 'What Comes
Beyond the Standard Models?' Bled, Slovenia, July 17-27, 2007
(DMFA, Zaloznistvo, Ljubljana, 2007); arXiv:0711.4681.
\item
L.V.~Laperashvili,\\ 
{\it The standard model and the fine structure constant at Planck distances in Bennet-Brene-Nielsen-Picek random dynamics,}\\
Phys. Atom. Nucl. {\bf 57}, 471 (1994)
[Yad. Fiz. {\bf 57}, 501 (1994)].
\seti
\end{enumerate}
\end{frame}

\begin{frame}
\frametitle{References}
\begin{enumerate}
\conti
\item
C.D.~Froggatt, L.V.~Laperashvili and H.B. Nielsen,\\
{\it The Fundamental-weak scale hierarchy in the Standard
Model,}\\ Phys. Atom. Nucl. {\bf 69}, 67 (2006), arXiv:hep-ph/0407102.
\item
C.D.~Froggatt and H.B.~Nielsen,\\ {\it Standard model criticality
prediction top mass 173 $\pm$ 5 GeV and Higgs mass 135 $\pm$ 9
GeV,}\\ Phys. Lett. B {\bf 368}, 96 (1996), arXiv:hep-ph/9511371.
\item
G.~Degrassi, S.~Di Vita, J.~Elias-Miro,
J.R.~Espinosa, G.F.~Giudice, G.~Isidori and A.~Strumia,\\
{\it Higgs mass and vacuum stability in the Standard Model at NNLO,}\\
JHEP {\bf 1208}, 098 (2012), arXiv:1205.6497.
\item
C.R.~Das, L.V.~Laperashvili, H.B.~Nielsen and
A.~Tureanu,\\
{\it Mirror world and superstring-inspired hidden sector of the
Universe, dark matter and dark energy,}\\
Phys. Rev. D {\bf 84}, 063510 (2011), arXiv:1101.4558.
\seti
\end{enumerate}
\end{frame}

\begin{frame}
\frametitle{References}
\begin{enumerate}
\conti
\item
C.R.~Das, L.V.~Laperashvili and
A.~Tureanu,\\
{\it Cosmological Constant in a Model with Superstring-Inspired
E(6) Unification and Shadow Theta-Particles,}\\
Eur. Phys. J. C {\bf 66}, 307 (2010), arXiv:0902.4874.
\item
D.L.~Bennett, L.V.~Laperashvili, H.B.~Nielsen and
A.~Tureanu,\\
{\it Gravity and mirror gravity in Plebanski formulation,}\\
Int. J. Mod. Phys. A {\bf 28}, 1350035 (2013), arXiv:1206.3497.
\seti
\end{enumerate}
\end{frame}

\end{document}